\documentclass{scrartcl}
\usepackage{bbm,amsfonts,amssymb,epsfig,verbatim,mathbbol}

\begin{document}
\title{\begin{flushright}
\normalsize
{\rm\large INLO-PUB-05/03}\\
{\rm\large FSU TPI 04/03}\\
\end{flushright}
\vspace{1cm}
A ring of instantons inducing a monopole loop}
\author{{\bf Falk Bruckmann} ${}^{a}$ and {\bf D\"orte Hansen} ${}^{b}$\\
\\
${}^{a}$ Instituut-Lorentz for Theoretical Physics, University of Leiden,\\
P.O.Box 9506, NL-2300 RA Leiden, The Netherlands\\
{\tt bruckmann@lorentz.leidenuniv.nl}\\
\\
${}^{b}$ Institute of Theoretical Physics, Friedrich-Schiller-University Jena,\\ 
Max-Wien-Platz 1, D-07743 Jena, Germany\\
{\tt nch@tpi.uni-jena.de}
}
\date{}
\maketitle

\begin{abstract}
We consider the superposition of infinitely many
instantons on a circle in $\mathbb{R}^4$. The construction yields a self-dual
solution of the Yang-Mills equations with action density concentrated on the
ring. We show that this configuration is reducible in which case magnetic charge
can be defined in a gauge invariant way. Indeed, we find a unit charge monopole
(worldline) on the ring. This is an analytic example of the correlation between
monopoles and action/topological density, however with infinite action. We show
that both the Maximal Abelian Gauge and the Laplacian Abelian Gauge detect the
monopole, while the Polyakov gauge does not. We discuss the implications of this
configuration.
\end{abstract}
 
\def\d{{\rm d}}
\def\D{{\rm D}}

\def\mR{\mathcal{R}}
\def\pa{\partial}
\def\be{\bar{\eta}}
\def\a{\alpha}
\def\b{\beta}
\def\eps{\epsilon}
\def\vp{\varphi}
\def\l{\lambda}
\def\vt{\vartheta}
\def\pl{\parallel}
\def\mb{\mathbf}

\def\Tr{{\rm{Tr}}}
\def\tu{\tilde{u}}
\def\tv{\tilde{v}}
\def\tF{\tilde{F}}
\def\ba{\bar{\eta}}
\def\m{\mathcal}
\def\S{\mathcal{S}}


\section{Introduction}

Quantum Chromodynamics as the theory of strong
interactions exhibits interesting non-perturbative phenomena at low energies, among
them chiral symmetry breaking and confinement.
While the first of them is explained by \textit{instanton} related mechanisms,
confinement has remained a puzzle and other topological excitations have
been proposed to explain this effect.
One of the most popular models is the dual superconductor scenario
\cite{thooft:76a,mandelstam:76}. In this picture, the
condensation of \textit{magnetic monopoles} leads to flux tubes
between chromoelectric charges. The fundamental problem of this
model is the fact that monopoles are a priori \textit{not} present
in Yang-Mills theories. Nevertheless, it is possible to introduce them
by partial gauge fixings called Abelian gauges (AG) \cite{thooft:81a}, 
where the gauge freedom is used to diagonalise
an auxiliary operator $O(x)$ transforming in the adjoint
representation. Then
monopoles occur as defects of the gauge fixing at points where
$O(x)$ has coinciding eigenvalues.

However, this approach suffers from the weakness that depending on the choice of
the operator $O$ one ends up with different Abelian gauges, the best studied
examples being the Maximally Abelian Gauge (MAG), the Laplacian Abelian
Gauge (LAG) and the Polyakov gauge (PG), respectively. It is 
still under discussion whether all AG's are equally useful or one of them is
best suited for solving the confinement problem.

There is a strong hint that instantons and the
configurations responsible for confinement are related:
Lattice simulations have shown that chiral symmetry breaking and confinement
have the same critical temperature \cite{kogut:83}. 
In other words, instantons and monopoles\footnote{and presumably also
vortices} may occur on the same footing.
Since these objects are both of topological origin, 
a relation between the monopole charge and the instanton number in
Abelian gauges has been derived
\cite{jahn:00}.
This may be seen as some gauge-independent content of Abelian gauges.

Moreover, the monopole position is often correlated to lumps
of the action/topological charge density.
The first observation in this direction
was the finding that the high temperature limit of the
Harrington-Shepard caloron \cite{harrington:78}
leads to a static BPS monopole \cite{rossi:79}, i.e.~a monopole
line. In the same spirit we expect a monopole \textit{loop} to be induced by a
configuration with topological charge concentrated on a ring.
Two overlapping instantons as well as a finite number of instantons are known
to possess this property \cite{garciaperez:00,brihaye:89}. Pushing this to the limit, our investigation
will concern the \textit{superposition of infinitely many instantons on a
ring}.
The gauge field constructed this way will come out to be
$U(1)$-\textit{reducible}, i.e.~effectively Abelian.
Therefore there is a natural
operator $O$ to diagonalise and this gauge transformation
in fact induces a monopole on the ring.
This singularity prevents the configuration from having finite
action. We will use it for kinematical aspects, namely
as a test for the mentioned popular Abelian gauges.
Throughout the paper we will restrict ourselves to gauge group $SU(2)$.

\section{Construction of the instanton ring}

The general instanton in four dimensional Euclidean space-time can be
obtained by the (algebraic and nonlinear) ADHM construction \cite{atiyah:78}, while
a subclass with restricted relative color orientation of the
constituents fits into the ansatz,
\begin{eqnarray}
A_\mu=A_\mu^a\:\frac{\sigma_a}{2}\,,\qquad
A_\mu^a=-\be^a_{\mu\nu}\pa_\nu\ln\Pi(x)\,,
\end{eqnarray}
where $\be$ and $\sigma$ are the 't Hooft tensor and the Pauli
matrices, respectively. Within this ansatz the
self-duality condition
$F_{\mu\nu}=\frac{1}{2}\eps_{\mu\nu\rho\sigma}F_{\rho\sigma}$  reduces to the
condition $\square\Pi/\Pi=0$ 
which is solved by the Green's function of point `charges' at positions
$y_k$ with scale parameters $\rho_k$.
We will use the 't Hooft ansatz \cite{tHooft:76d}
and conformal ansatz \cite{jackiw:77},
in which the scalar potential $\Pi$ for a charge $N$
instanton reads
\begin{eqnarray}
\Pi(x)= 1+ \sum_{k=1}^N\frac{\rho_k^2}{(x-y_k)^2}\,,\qquad
\Pi(x)=\sum_{k=1}^{N+1}\frac{\rho_k^2}{(x-y_k)^2}\,,\label{ansaetze}
\end{eqnarray}
respectively.
In both cases, the topological charge density 
can be written quite simply as
$\Tr \, F_{\mu\nu}F_{\mu\nu}=\square\square\ln \Pi\,$.

The superposition of infinitely many
BPST instantons along the time axis leads to the Harrington-Shepard caloron
\cite{harrington:78} which at high temperature turns
into the BPS monopole \cite{rossi:79}. In the limit of vanishing distance between
the instanton copies this construction
amounts to the following potential,
\begin{eqnarray}
\Pi(x)= \int_{-\infty}^\infty \d t\:\frac{\rho^2}{\mb{x}^2
     +(x_4-t)^2} =\frac{\pi\rho^2}{|\mb{x}|}\,.
\end{eqnarray}

Inspired by this construction we consider now the configuration of 
\textit{infinitely many instantons with same size equidistantly placed
on a circle}. We choose this circle to be in the $x_1x_2$-plane,
\begin{eqnarray}
y_{k\,\mu}=R\,(\cos(2\pi k/N),\sin(2\pi k/N),0,0)\,,\qquad \rho_k=\rho\,.
\end{eqnarray}
An arbitrary location of the circle
can be gained by the action of the symmetry group SO(4).
For our calculations, double polar coordinates
\begin{eqnarray}
      x_\mu=(r_{12}\cos\vp_{12},\, r_{12}\sin\vp_{12},\,
      r_{34}\cos\vp_{34},\, r_{34}\sin\vp_{34})
\end{eqnarray}
are most suitable. The limit $N\rightarrow\infty$ can be performed
as the integral
\begin{eqnarray}
\Pi(x)=a+\frac{N\rho^2}{2\pi}\int_0^{2\pi}
      \!\!\d\xi\:\frac{1}{r_{12}^2+R^2-2Rr_{12}\cos(\vp_{12}-\xi)+r_{34}^2}\,,\qquad
      \xi\sim\frac{2\pi k}{N}\,.
\end{eqnarray}
In this limit, the only difference between the two ans\"atze (\ref{ansaetze}) is
the first term which we parametrise by $a=1$ for the 't Hooft ansatz and $a=0$
for the conformal ansatz, respectively.
In order to get a well behaved $\Pi$, the size $\rho^2$ has to decrease
to zero when taking $N$ to infinity\footnote{Allowing $N\rho^2$ to diverge
automatically leads to the conformal ansatz, since then $a$ can be neglected
(and the divergence is multiplicative).}, that is we perform the limit
\begin{eqnarray}
      N\rho^2\stackrel{N\to\infty}{\longrightarrow}\mbox{const}\equiv\l^2.
\end{eqnarray}
Then the integral yields,
\begin{eqnarray}
\Pi(x)=a+\frac{\l^2}{S(r_{12},r_{34})}\,,\qquad
S(r_{12},r_{34})=\sqrt{(r_{12}^2+r_{34}^2-R^2)^2+4r_{34}^2R^2}\,.
\label{phi_ring}
\end{eqnarray}
In \cite{brihaye:89} this formula was found by summability of the
finite $N$ case. Aspects of such superpositions from the viewpoint of hyperbolic
monopoles have been studied in \cite{chakrabarti:86}.

$S(r_{12},r_{34})$ vanishes at the circle $r_{12}=R,\,r_{34}=0$. Therefore, $\Pi$ is singular there.
Not surprisingly the latter is the Greens
function of the Laplacian with a constant charge density on the ring.
The ring ansatz together with the limit $N\rightarrow \infty$ has lead to a symmetry
$SO(2)\times SO(2)$ since $\Pi$ only depends on the two radii.

We note in passing that the scalar potential (\ref{phi_ring}) can be obtained
from the one of the single instanton by complexifying the radius $r_{34}$ and
computing the absolute value, $\Pi(x)=a+\lambda^2/|r_{12}^2+(r_{34}\pm iR)^2|$.

Seen from far away, the potential behaves as
\begin{eqnarray*}
      \Pi(x)=a+\frac{\l^2}{r^2},\qquad r\equiv\sqrt{r_{12}^2+r_{34}^2}\gg R.
\end{eqnarray*}
Thus the ring configuration looks like the ordinary BPST instanton
in singular gauge for $a=1$  and like pure gauge for $a=0$, respectively.
However, the topological charge of these configurations is
infinite by construction. The lumpiness of the action density
manifests itself in an extreme way, namely as a non-integrable
singularity at the ring where both ans\"atze behave as
\begin{eqnarray}
\Tr\, F_{\mu\nu}F_{\mu\nu}=-32 \,\frac{R^4}{S(r_{12},r_{34})^4}
\qquad\mbox{as }r_{12}\rightarrow R,\,r_{34}\rightarrow 0
\end{eqnarray}
We will comment more on this singularity later.

\section{Reducibility and monopole content}

It is straightforward to calculate the field strength of the ring
configuration (\ref{phi_ring}).
Since self-duality is still fulfilled,
the chromoelectric field is sufficient.
For the conformal case $a=0$ one finds,
\begin{eqnarray}
E_i^a(x)=f_i(x)n^a(x)\,.\label{electric_field}
\end{eqnarray}
This means that \textit{all components of the field strength point in the same
color direction}.
This can be made explicit\footnote{DH
thanks M. Garcia--Perez for marking this point.} by the vanishing of the matrix
$M_{ij}=(E_i^a)^2(E_j^b)^2-(E_i^aE_j^a)^2$.
The direction in color space is given by the normalised vector field $n$,
\begin{eqnarray}
n^a(x)=\frac{1}{S(r_{12},r_{34})}\left(\begin{array} {c}
       2\,(-x_1x_3-x_2x_4) \\
       2\,(x_1x_4-x_2x_3) \\
       r_{12}^2-r_{34}^2-R^2 \end{array}\right).\label{n_field}
\end{eqnarray}

\begin{figure}[!b]
\begin{center}
\epsfig{figure=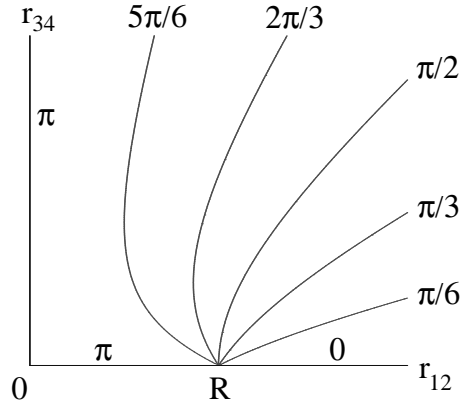,scale=0.4}
\end{center}
\caption{Lines of constant azimuthal angle $\beta$ for
    the normalised Higgs field $n$ as given by $n^{a=3}(x)=\cos\beta(x)$ and (\ref{n_field}),
    basically the same as in \cite{brower:97b,jahn:00}.}
\label{n_figure}
\end{figure}

Such a behaviour is a strong hint for the \textit{reducibility} of the
configuration. Loosely speaking, the configuration is Abelian with a
space-dependent embedding $n^a(x)$ of the Abelian direction into the
non-Abelian color space. More precisely, the definition of
reducibility involves a normalised `Higgs' field $n$ in the adjoint
representation being
covariantly conserved in the
background of $A$,
\begin{eqnarray}
      D_\mu n\equiv \pa_\mu n-i[A_\mu,n]\equiv 0\,.\label{cov_const}
\end{eqnarray}
A related characterisation of reducibility is the non-trivial stabiliser of $A$,
namely the $U(1)$ subgroup $\exp(i\mu n(x))$ with constant $\mu$.
From the gauge covariance of (\ref{cov_const}) follows that
on diagonalising $n$, the gauge field $A$ becomes purely diagonal,
i.e.~Abelian. Accordingly, all components of the field strength become
diagonal. This or equivalently
the integrability condition $[D_\mu,D_\nu]n=[F_{\mu\nu},n]=0$
gives the particular
behaviour of the field strength mentioned above.
All this can be verified for the ring in the conformal ansatz, but does not hold
for the 't Hooft ansatz.
This indicates that the relative orientation of the constituents -- which in
general is different for the two ans\"atze as can be seen from the completeness
of the conformal ansatz for topological charge 2 -- is important 
for properties like reducibility and the monopole content, which we
discuss in a moment. From now on we will consider only the conformal case $a=0$.

The $f_i$'s are scalar functions of a similar structure as $n$,
\begin{eqnarray}
f_i(x)=\frac{4\,R^2}{S(r_{12},r_{34})^3}\left(\begin{array} {c}
       2\,(-x_1x_3+x_2x_4) \\
       2\,(x_1x_4+x_2x_3) \\
       r_{12}^2-r_{34}^2-R^2 \end{array}\right),
\end{eqnarray}
which is typical for symmetric configurations.

Reducible gauge fields can be decomposed in a natural way w.r.t.~$n$,
\begin{eqnarray}
A_\mu^a=C_\mu n^a-\eps_{abc}n^b\pa_\mu n^c\,,\qquad
F_{\mu\nu}^a=(\epsilon_{bcd}n^b\pa_\mu
n^c\pa_\nu n^d+\pa_\mu C_\nu-\pa_\nu C_\mu)n^a
\,,\label{cho_conn}
\end{eqnarray}
known as
Cho connections \cite{cho:80}.
$C_\mu=A_\mu^a n^a$ plays the role of an Abelian gauge field,
for the ring it can be written in a compact form as
\begin{eqnarray}
C_\mu=\frac{2}{S(r_{12},r_{34})}\,(x_2,-x_1,x_4,-x_3)
=2\eta^3_{\mu\nu}\frac{x_\nu}{S(r_{12},r_{34})}\,.
\end{eqnarray}

Coming back to our original motivation we now investigate whether the
ring configuration has a monopole content.
Reducibility of the ring configuration provides a natural operator to
diagonalise in the spirit of Abelian gauges, namely $n$.
Let us repeat that $n$ can be obtained from such a particular
gauge field via the field strength (\ref{electric_field})
or the conservation equation (\ref{cov_const}).
The normalised field $n$ takes values in $S^2\subset su(2)$ which is
exactly the coset $SU(2)/U(1)$ fixed by Abelian gauges.

For further computations it is helpful to parametrise $n$ in terms of
Euler angles, $n^a(x)=
       (\sin\b(x)\,\cos\a(x),\,
       \sin\b(x)\,\sin\a(x),\,
       \cos\b(x))^T$.
In Fig.~\ref{n_figure} we depict the azimuthal angle $\beta(x)$.
The polar angle $\a$ is simply given by
$\a(x)=\vp_{12}-\vp_{34}+\pi$.
From this it is easy to see that the $n$-field
is a \textit{hedgehog} around the ring (for fixed
worldline variable $\varphi_{12}$), i.e.~in the vicinity of the ring
it takes on all values once. It behaves like the asymptotic Higgs field in
the 't Hooft-Polyakov monopole. Thus, the diagonalisation\footnote{which means
going to the unitary gauge} of $n$ introduces a \textit{unit charge Dirac
monopole with its worldline being the ring}.

\section{Detection of the monopole in Abelian gauges}

With the help of reducibility we have shown that the ring
configuration contains a monopole loop in a canonical way.
In this section we will
test
whether the three most popular Abelian gauges recognize the monopole, too.

\subsection{Maximally Abelian gauge}

In the continuum, the MAG is defined by the gauge condition
\begin{eqnarray}
\pa_\mu A_\mu^\perp-i[A_\mu^3,A_\mu^\perp]=0\,,\label{MAG_diff}
\end{eqnarray}
where $A_\mu^\perp=A_\mu^1\sigma_1/2+A_\mu^2\sigma_2/2$ represents the off-diagonal
gauge field \cite{thooft:81a}. Using the properties of the $\bar{\eta}$-symbol, it is
clear that the instanton ring both in the conformal and 't Hooft
ansatz satisfies the MAG condition.

In order to come to a unique gauge fixing
the above condition has to be restricted
further. The gauge condition is endowed by the minimium of the
following MAG functional,
\begin{eqnarray}
F_{\rm MAG}[A_\mu]=\frac{1}{2}\int \d^4x\: \Tr
(A_\mu^\perp)^2\ \to \rm{min}\,.\label{MAG_func}
\end{eqnarray}
This can be read as going from a solution of the equation of motion (\ref{MAG_diff}) to an
actual minimum of the action (\ref{MAG_func}).

The singularity at the ring does not spoil the integrability of this
functional for the configuration at hand. It turns out that the
functional diverges due to its behaviour at infinity,
$F_{\rm MAG}[A_\mu]=\int\d^4 x\, r^2/S^2=\infty$.
On the other hand, we know from reducibility that there exists a gauge transformation
which makes $A_\mu$ purely diagonal. Accordingly, the MAG-functional
takes its smallest possible value, namely zero.
This is a general property of reducible configurations/Cho connections
(\ref{cho_conn}).

Observe that the singular gauge transformation 
bringing the configuration into the MAG is exactly the one which
introduces the Dirac monopole.
This is supported by an equivalent version of the MAG which defines
its (normalised) Higgs field $n$ as the one which minimizes
$\int\d^4 x\,\Tr(\D_\mu n)^2$. Indeed this alternative MAG-functional vanishes
for the $n$-field of the ring due to equation (\ref{cov_const}). 
We conclude that the Maximal Abelian
gauge indeed \textit{recognizes the monopole content of this configuration.}
However, it is not sensitive to the size of the ring since the
functional becomes zero independent of the radius $R$.
Actually, this statement also holds for the 't Hooft ansatz, where the MAG-functional is
finite from the beginning. 
These findings should be compared to the fact that the monopole loop in the
background of a single instanton is suppressed by the MAG-functional
\cite{brower:97b}.  

\subsection{Laplacian Abelian gauge}

The auxiliary operator $O(x)$ of the LAG is the ground state $\phi$ of the
gauge covariant Laplacian in the adjoint representation \cite{vandersijs:97},
\begin{eqnarray}
-D_\mu^2\phi=E_0\phi\,.
\end{eqnarray}
Thus this field has to be square integrable. The $n$-field of the ring
is a zero mode of the Laplacian due to (\ref{cov_const}), but being
pointwise normalised, $n^a(x)n^a(x)=1$, it is clearly not square
integrable. This is a known problem of this gauge when defined over infinite
volume manifolds. In order to circumvent it, conformal invariance was
used to go onto the four-sphere \cite{bruckmann:01a}. Since the
covariant Laplacian there reads
\begin{eqnarray}
-D^2=-\frac{1}{\sqrt{g}}D_\mu\sqrt{g}g^{\mu\nu}D_\nu\,,
\end{eqnarray}
$n$ is again a zero mode due to the same property (\ref{cov_const}).
Moreover, $n$ is now square integrable since the problem at infinity
is gone. Thus $n$ is the operator of the LAG and therefore 
the LAG \textit{detects the monopole loop}, too.

\subsection{Polyakov gauge}

The Polyakov gauge is adapted to finite temperature or other
situations where at least one of the coordinates is compact. The
operator $O(x)$ in this case is the Wilson loop in this `time' direction,
\begin{eqnarray}
\mathcal{P}(\vec{x})
=\mathcal{P}\exp\,(\,i\!\!\int_0^{T}\!\!\!A_4(x_4,\vec{x})\:\d x_4)\,.
\end{eqnarray}
It is a group valued variable, and for gauge group $SU(2)$ the defects
occur at $\mathcal{P}(\vec{x})=\pm\Eins$, i.e.~tr$\,\mathcal{P}(\vec{x})=\pm 2$
\cite{reinhardt:97b,ford:98,jahn:98}.
The diagonalising transformation of the Polyakov loop is known explicitly
and can be
understood as coming close to the Weyl gauge $A_4=0$ while respecting
the holonomy in this direction.

The Polyakov gauge can also be applied to the single instanton
on four-dimensional Euclidean space by virtue of the fact that the according
unlimited integral in the Polyakov line
\begin{eqnarray}
\mathcal{P}(\vec{x})
=\mathcal{P}\exp\,(\,i\!\!\int_{-\infty}^{\infty}\!\!\!A_4(x_4,\vec{x})\:\d x_4)
\end{eqnarray}
exists.
The instanton reveals a monopole which is static (by definition), runs
through the instanton position and shows a perfect hedgehog behaviour
in the spacial directions.

A priori it is not obvious, whether the Polyakov gauge is applicable to the
ring in the same spirit. There are essentially two different choices of a `time'
direction, $x_4$ and $x_1$. In the first case a whole monopole loop
appears at some instant in `time'. In the second case there is a monopole pair
creation, a loop `evolution' and finally 
the annihilation of the pair.
The calculation of the Polyakov line in both cases does not involve
path ordering when performed at the monopole position.
As a matter of fact, the Polyakov line in the
first case is traceless, so does not see the monopole as a defect. In the second case one has to compute the
following integral
\begin{eqnarray}
\mbox{tr}\,\mathcal{P}(\vec{x})=2\cos\,(\!\int_{-\infty}^\infty\frac{x_2}{x_1^2+x_2^2-R^2}\:\d
x_1)\,.
\end{eqnarray}
For the range of interest $x_2\in[-R,R]$ the Cauchy principal value of this
integral gives indeed tr$\,\mathcal{P}=2$, but in addition there are infinitely
many defects just outside the monopole since the integral diverges as
$|x_2|\rightarrow R+0$. Altogether the Polyakov gauge \textit{is not appropriate} in our situation.

\section{Discussion}

We have constructed a gauge field coming from a superposition of
infinitely many instantons on a ring. As an overlap effect the ring becomes the
worldline of a magnetic monopole. This is a gauge-independent statement since the
configuration at hand is $U(1)$-reducible, i.e.~essentially Abelian. It results in all
components of the electric and magnetic field being parallel to each other in color
space. Put differently, the configuration is an example of a Cho connection
with non-trivial fields $n$ and $C$.

We have studied the application of Maximally Abelian gauge, Laplacian
Abelian gauge and Polyakov gauge to this configuration. It turns out
that the first two of them do recognize the monopole loop. This
fact is merely based on reducibility. The Polyakov gauge fails to do
so, which we interpret as an
effect of the superposition; the Polyakov gauge is questionable in this
situation anyhow.

Of particular interest is the MAG-functional since its minimisation is as difficult
as a spinglass problem. We propose to study the scalar potential
$\Pi(x) = 1+\rho^2/r^2 +\l^2/S(r_{12},r_{34})$, that corresponds to a
combination of an instanton (in singular gauge) and a ring 
configuration at radius $R$.
A minimum for $F_{\rm MAG}$, in particular for non-trivial values
of $\rho$, $\lambda$ and $R$, could give more insight into how the functional
weights instantons and monopoles and how they influence each other.

The ring configuration is an analytic example that monopoles come with an excess
of topological density as has been observed on the lattice. However,
by construction the ring has infinite
action\footnote{there are no strange effects when superposing the
constituents, which would make the action somehow finite}.
This should be
compared to the BPS monopole\footnote{actually a conformal
transformation connects the two}
(better the dyon) which has finite energy
and is static; therefore its action for an infinite time extension
becomes divergent. Since in our case the analogue of time -- the angle
$\varphi_{12}$ -- is compact, the divergence has to be present
already for fixed `$\varphi_{12}$-slice' near the ring.
Infact, the divergence of the action comes from a non-integrable singularity at
the ring: $S(r_{12},r_{34})$ is proportional to the distance
$\bar{r}=\sqrt{(r_{12}-R)^2+r_{34}^2}$ from the ring and the action density
behaves as $1/\bar{r}^4$ which cannot be cured by the measure.

This pathological behaviour is provoked by the complete
reducibility of the configuration. A Cho connection around the non-continuity of
$n$ always yields $A\sim\d n\sim 1/\bar{r}$ which results in the described
divergence for the action density. We have also investigated the possibility of
keeping reducibility but relaxing selfduality, that is keeping $n$ but
deforming $C$ say near the ring. It is possible to soften the singularity in the
topological density to an integrable one. However, the action density stays infinite
because its leading singularity comes from $\d n$ alone.

Therefore a more realistic configuration is expected to be Abelian (reducible)
only outside a non-Abelian core. Such a mechanism is at work for calorons with
non-trivial holonomy \cite{lee:98,kraan:98a}. Since these are built from a superposition of relatively
oriented instantons, one has to use the ADHM construction. The latter will make
the ring superposition much more complicated. Observe that also in this situation the alternative
MAG-functional will be close to zero, since its integrand $\D_\mu n$ vanishes in
a large fraction of space-time.

The relative color orientation of constituents has been shown to be crucial for the instanton approach to
confinement \cite{hart:96}. Our findings that the
conformal ansatz induces a monopole loop while the 't Hooft ansatz does not,
supports this fact. More knowledge is needed of how to build instanton
clusters or chains in order to obtain a monopole worldline. The percolation of a
monopole loop through the volume -- as one criterion for confinement -- should
then be induced by an instanton ensemble.

\section{Acknowledgements}

The authors are grateful to Chris Ford, Tom Heinzl, Pierre van Baal and Andreas
Wipf for helpful discussions.
DH thanks M. Garcia-Perez for the hospitality at CERN and
for clarifying some points.
The research of FB is supported by FOM.

\begingroup\raggedright\endgroup


\begin{thebibliography}{10}

\bibitem{thooft:76a}
G.~'t~Hooft. in: \textit{High Energy Physics}, Proceedings of the EPS
  International Conference, Palermo 1975, A.~Zichichi, ed., Editrice
  Compositori, Bologna 1976.

\bibitem{mandelstam:76}
S.~Mandelstam, {\it Vortices and quark confinement in non-Abelian gauge
  theories},  {\em Phys.~Rep.} {\bf C23} (1976) 245--249.

\bibitem{thooft:81a}
G.~'t~Hooft, {\it Topology of the gauge condition and new confinement phases in
  non-Abelian gauge theories},  {\em Nucl.~Phys.} {\bf B190} (1981) 455.

\bibitem{kogut:83}
J.~Kogut, M.~Stone, H.~W. Wyld, W.~R. Gibbs, J.~Shigemitsu, S.~H. Shenker, and
  D.~K. Sinclair, {\it Deconfinement and chiral symmetry restoration at finite
  temperature in SU(2) and SU(3) gauge theories},  {\em Phys.~Rev.~Lett.} {\bf
  50} (1983) 393.

\bibitem{jahn:00}
O.~Jahn, {\it Instantons and monopoles in general Abelian gauges},  {\em
  J.~Phys.} {\bf A33} (2000) 2997--3019,
  {\tt hep-th/9909004}.

\bibitem{harrington:78}
B.~J. Harrington and H.~K. Shepard, {\it Periodic Euclidean solutions and the
  finite temperature Yang-Mills gas},  {\em Phys.~Rev.} {\bf D17} (1978) 2122.

\bibitem{rossi:79}
P.~Rossi, {\it Propagation functions in the field of a monopole},  {\em
  Nucl.~Phys.} {\bf B149} (1979) 170.

\bibitem{garciaperez:00}
M.~{Garcia~Perez}, T.~G. Kovacs, and P.~van Baal, {\it Overlapping instantons},
   {\tt hep-ph/0006155}.

\bibitem{brihaye:89}
Y.~Brihaye and J.~Kunz, {\it Summable chains of instantons and their
  symmetries},  {\em J.~Math.~Phys.} {\bf 30} (1989) 1913--1917.

\bibitem{atiyah:78}
M.~F. Atiyah, N.~J. Hitchin, V.~G. Drinfeld, and Y.~A. Manin, {\it Construction
  of instantons},  {\em Phys.~Lett.} {\bf A65} (1978) 185--187.

\bibitem{tHooft:76d}
G.~'t~Hooft {\em unpublished} (1976).

\bibitem{jackiw:77}
R.~Jackiw, C.~Nohl, and C.~Rebbi, {\it Conformal properties of pseudoparticle
  configurations},  {\em Phys.~Rev.} {\bf D15} (1977) 1642--1646.

\bibitem{chakrabarti:86}
A.~Chakrabarti, {\it Construction of hyperbolic monopoles},  {\em
  J.~Math.~Phys.} {\bf 27} (1986) 340--348.

\bibitem{brower:97b}
R.~C. Brower, K.~N. Orginos, and C.-I. Tan, {\it Magnetic monopole loop for the
  Yang-Mills instanton},  {\em Phys.~Rev.} {\bf D55} (1997) 6313,
  {\tt hep-th/9610101}.

\bibitem{cho:80}
Y.~M. Cho, {\it A restricted gauge theory},  {\em Phys.~Rev.} {\bf D21} (1980)
  1080.

\bibitem{vandersijs:97}
A.~J. van~der Sijs, {\it Laplacian Abelian projection},  {\em Nucl.~Phys.~B
  (Proc.~Suppl.)} {\bf 53} (1997) 535,
  {\tt hep-lat/9608041}.

\bibitem{bruckmann:01a}
F.~Bruckmann, T.~Heinzl, T.~Vekua, and A.~Wipf, {\it Magnetic monopoles
  vs.~Hopf defects in the Laplacian (Abelian) gauge},  {\em Nucl.~Phys.} {\bf
  B593} (2001) 545--561, {\tt hep-th/0007119}.

\bibitem{reinhardt:97b}
H.~Reinhardt, {\it Resolution of Gauss' law in Yang-Mills theory by gauge
  invariant projection: Topology and magnetic monopoles},  {\em Nucl.~Phys.}
  {\bf B503} (1997) 505--529, {\tt hep-th/9702049}.

\bibitem{ford:98}
C.~Ford, U.~G. Mitreuter, J.~M. Pawlowski, T.~Tok, and A.~Wipf, {\it Monopoles,
  Polyakov loops and gauge fixing on the torus},  {\em Ann.~Phys.~(N.Y.)} {\bf
  269} (1998) 26, {\tt hep-th/9802191}.

\bibitem{jahn:98}
O.~Jahn and F.~Lenz, {\it Structure and dynamics of monopoles in axial gauge
  QCD},  {\em Phys.~Rev.} {\bf D58} (1998) 085006,
  {\tt hep-th/9803177}



\bibitem{lee:98}
K.~Lee and C.~Lu, {\it SU(2) calorons and magnetic monopoles},  {\em
  Phys.~Rev.} {\bf D58} (1998) 025011,
  {\tt hep-th/9802108}.

\bibitem{kraan:98a}
T.~C. Kraan and P.~van Baal, {\it Periodic instantons with non-trivial
  holonomy},  {\em Nucl.~Phys.} {\bf B533} (1998) 627--659,
  {\tt hep-th/9805168}.

\bibitem{hart:96}
A.~Hart and M.~Teper, {\it Instantons and monopoles in the maximally Abelian
  gauge},  {\em Phys.~Lett.} {\bf B371} (1996) 261--269,
  {\tt hep-lat/9511016}.

\end{thebibliography}
\end{document}